\documentclass[sigconf]{acmart}

\usepackage{epsfig}
\usepackage{xcolor, colortbl}
\definecolor{citegreen}{HTML}{458B00}
\usepackage{adjustbox}
\usepackage{array}
\usepackage{lipsum}
\usepackage{enumitem} % for changing enumerate to letters
\begin{document}

\title{Unsupervised Learning for security of Enterprise networks by micro-segmentation}

\author{Mahmood Yousefi-Azar}
\affiliation{Macquarie University, Sydney}
%\email{mahmood.yousefiazar@mq.edu.au}
\email{mahmood@ditno.com}

\author{Mohamed Ali Kaafar}
\affiliation{
%Faculty of Science and Engineering \\
Macquarie University, Sydney}
\email{dali.kaafar@mq.edu.au}

\author{Andy Walker}
\affiliation{Ditno. Pty Ltd, Sydney, Australia}
\email{andy@ditno.com}

\begin{abstract}

Micro-segmentation as a network security technique requires delivering services for each unique segment. To do so, the first stage is defining these unique segments (a.k.a security groups) and then initializing policy-driven security controls. In this paper, we propose an unsupervised learning technique that covers both the security grouping and policy creation. For the network asset grouping, we develop a distance-based machine learning algorithm using the dynamic behavior of the assets. That is, after observing the entire network logs, our unsupervised learning algorithm suggests partitioning network assets into the groups. A key point of this unsupervised technique is that the grouping is only generated during the training phase and remains valid during the testing phase. The outcome of the grouping stage is then fed into the rules (security policies) creation stage enabling to establish the security groups as the lowest granularity of a firewall rules.  

We conducted both quantitative and qualitative experiments and demonstrate the good performance of our network micro-segmentation approach. We further developed a prototype to validate the run-time performance of our approach at scale in real-world environment. The hyper-parameters of our approach provides users with a flexible model to be fine tuned to adapt very easily with the enterprise’s security governance.

\end{abstract}

\vspace{-0.1in} 


\maketitle

\vspace{-0.1in} 

\section{Introduction}

One of the critical and most important parts of network automation is the network security configuration. To define and enforce the security settings, consistently over an entire network, micro-segmentation\footnote{\url{https://www.networkworld.com/article/3247672/what-is-microsegmentation-how-getting-granular-improves-network-security.html}} helps to securely isolate network assets (referred here as endpoints) from each other. This not only reduces the surface of attacks but also enables to confine possible attacks by limiting their impact inside the network.

This paper focuses on two main parts of micro-segmentation: security grouping and firewall rules creation. The first step is grouping the endpoints of a network. That is, endpoints that must be enforced within  certain security policies, must be member of a given group. A traditional and naive approach for this task is network connectivity analysis and manually labeling and selection of each endpoint to belong to a group \cite{celebi2019distributed}. These approaches are however prone to human errors and misinterpretations leading to high security risks. They are also highly expensive, time consuming or altogether faulty. 

The second step is firewall rules creation, a process of building package filters by which the communication between two sides is either blocked or allowed. The main challenges of building these filters are completeness and compactness while it is critical to have them all consistent. To have such characteristics, it is common to construct a tree-shape decision model out of services components and then have a generalization phase followed by an anomaly removing step \cite{golnabi2006analysis}.

Machine learning techniques have been successfully used in a wide range of applications including computer network security \cite{shen2018tiresias,guo2018lemna}. In this work, we propose the use of an unsupervised machine learning algorithm to enable the aggregation of each endpoint dynamic behavior during a certain period of time for the purpose of security analysis. 

We set the dynamic behavior of endpoints as the feature space of the learning algorithm. After observing enough samples from a given endpoint, the algorithm suggests the endpoint’s likelihood of belonging to each security groups. The Rules creation step aims then at identifying legitimate communications between these groups.

We collected two datasets from Ditno\footnote{\url{https://www.ditno.com/}} and performed real world experimentations with both quantitative and qualitative tests conducted in a client environment. The contribution of this paper are as follows:

\begin{itemize}

\item We propose the application of an unsupervised machine learning algorithm for the grouping of endpoints for the micro-segmentation purpose. To the best of our knowledge, this is the first time unsupervised machine learning techniques are used for network micro-segmentation.

\item We propose an integrated model that captures endpoints' logs and provides a classification into security groups and automated firewall rules generation. 

\item We develop a highly efficient and scalable prototype for real world network applications.

\end{itemize}

\vspace{-0.1in} 

\section{The Proposed Scheme}

%\begin{figure*}
%  \centering
%  \includegraphics[width=0.8\linewidth]{Segmentation.png}
%  \caption{The high level fig.}
%\end{figure*}\label{fig:Segmentation}

\begin{figure*} [ht!]
  \centering
  \includegraphics[width=0.65\linewidth]{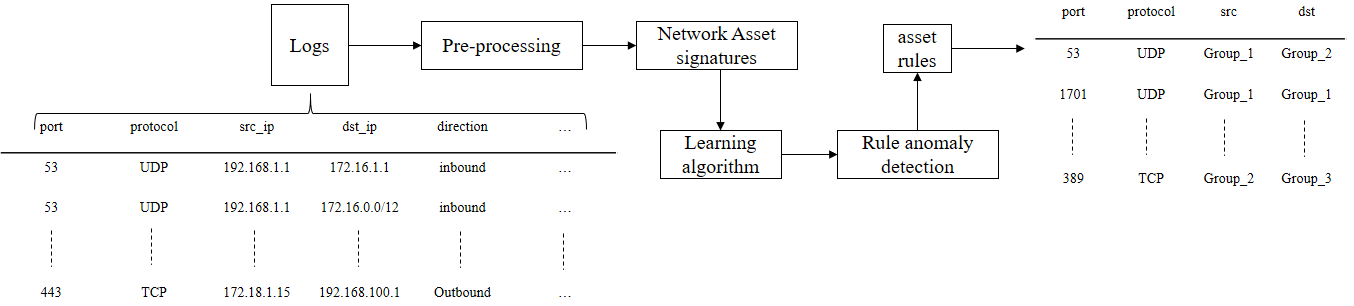}
  \caption{The proposed scheme. The values and abbreviations in the figure are only examples.\vspace{-0.1in} }
\label{fig:schemeMe}
\end{figure*}

\begin{table*} [ht!]
  \caption{The results of grouping phase excluding unknown communication side.\vspace{-0.1in} }
  \label{tab:resultsPure}
  \begin{tabular}{cccccccc}
    \toprule
    Dataset & Asset Qty & Group Qty &  Suggested group Qty & run-time  & homogeneity  & completeness & V-Measure \\
    \midrule
    Dataset{\tiny {1}} & 312 & 108 & 175 & 30.0s  & 98.24\%  & 84.76\% & 91.07\% \\
    Dataset{\tiny {2}} & 232 & 113  & 220 & 34.0s  & 96.13\%  & 80.42\%  &  87.56\% \\
    \bottomrule
  \end{tabular}
\end{table*}

%\vspace{-0.1in}

%Figure\ref{fig:schemeMe} shows a detailed schematic of the technique. Input data for the security grouping part is the real-time logs for each endpoint. Since the attributes of endpoints’ behavior are both numerical and categorical %For example, protocols are either TCP or UDP or ICMP while the frequency of unique traffic package for each communication is a real value.
%, to represent categorical attributes into real value format, we used one-hot encoding. The second part of pre-processing is removing unknown traffic logged from communication between network members with non-members. Also, it is also important to remove low frequency traffics since they influence endpoints’ main patterns. 

Figure\ref{fig:schemeMe} shows a detailed schematic of the proposed technique for micro-segmentation. Input data for the security grouping part is the real-time logs for each endpoint. Since the attributes of endpoints’ behavior are both numerical (e.g. the frequency of unique traffic package) and categorical (e.g. protocols), we used one-hot encoding to represent categorical attributes into real value format. The second part of pre-processing is removing unknown traffic logged from communication between network members with non-members. %Also, it is also important to remove low frequency traffics since they influence endpoints’ main patterns. 

After pre-processing phase, the signature of each endpoint is extracted. This signature can be extracted using a linear function (e.g. Principal Component Analysis) or non-linear function (e.g. Restricted Boltzmann Machine) \cite{cui2018survey}. These feature representation algorithms also reduce the feature space dimensionality and drastically reduce computing overhead that is vital for large scale networks. For the learning algorithm, we use K-means clustering algorithm as a distance-based metric. A key issue is that the number of security groups and thereby the number of clusters is not known beforehand. To address this, we simply assume a maximum number of potential security groups to be the number of endpoints. Because endpoints traffic type varies (i.e. the number of samples for each endpoint) depending on the endpoints activity, final group membership is obtained using the average distance from all the centroids.

To build the rules, the unique endpoints’ traffics are selected using the suggested group identifier. This represents an initial raw rule. To reduce the number of rules, it is common to generalize rules and then remove the rules that are known as anomaly \cite{golnabi2006analysis}. However, our rule generalization and anomaly detection are implemented in one step by replacing each endpoint IP with a corresponding network object. 

%To resolve rules with conflicts, we remove lower coverage objects. 

\vspace{.1 in}

\begin{table} [ht!]
  \caption{The results of the grouping phase including the network objects inclusion.\vspace{-0.1in} }
  \label{tab:resultsIP}
  \begin{tabular}{cccc}
    \toprule
    Dataset &  homogeneity  & completeness & V-Measure \\
    \midrule
    Dataset{\tiny {1}} & 99.84\% & 82.56\% &   90.37\%  \\
    Dataset{\tiny {2}} & 97.11\% & 81.18\%  &  88.41\% \\
    \bottomrule
  \end{tabular}
\end{table}%\label{tab:resultsIP}

\vspace{-0.1in} 

\section{Results}

To evaluate our system, we consider the metrics of homogeneity, completeness and V-measure as three key indicators of the quality of clustering operations \footnote{Readers may refer to \url{https://scikit-learn.org/stable/modules/clustering.html}
 for further details about these metrics)}. Table~\ref{tab:resultsPure} presents the results of the grouping phase compared with ground-truth. It is worth mentioning that hyper-parameters of the system are fixed to have a desirable homogeneity metric (e.g. $\geq$ 95\%) as a main concern from a security perspective. This is the reason that the learning algorithm suggests 175 and 220 security groups for the first and second datasets respectively while completeness is more than 80\%. V-measure, around 90\%, as the harmonic mean of homogeneity and completeness is our main metric.

In practice, there can be endpoints communicating with assets of the network while the other side of the communication is not logged. To address this problem, we replace the other side with network objects. Table~\ref{tab:resultsIP}  shows the results. The homogeneity increases for both datasets while the V-measure has slight improvement. 
We also had a qualitative test based on Ditno’s expertise and the outcome is satisfactory. Since grouping phase comes out of training phase, for new endpoints, the models is trained again using added new endpoints’ samples to data. In case this new data results in new clusters, the new endpoint establishes a new security group while not changing the number of clusters.

Figure\ref{fig:schemeMe} also presents some naive examples of the created rules. Since the rules are based on security groups suggested by the k-mean algorithm, only members of two groups can communicate in a given network service. This strategy resolves rules conflicts and redundancy. In particular, as classic rules generalization techniques are vulnerable to building any-to-any allowance problem, these groups based rules prevent this problem in the rule initiation phase. 

\vspace{-0.1in} 
\section{Conclusion}

This paper proposed a machine learning-based scheme for network micro-segmentation. We addressed two main problems: network security grouping and rule creation. We practically showed that a distance-based clustering algorithm such as k-means is high capability to provide users with security groups (i.e. clusters). These groups are obtained in the k-means training phase and using dynamic analysis of endpoints and the average distance to each cluster. This grouping technique is the basis of firewall rules where the lowest granularity of rules are the suggested groups rather than each target endpoint.

%\section{References}

\bibliographystyle{plain}
\bibliography{References} \label{References}

\end{document}